# Experimental Confirmation of the Universal Law for the Vibrational Density of States of Liquids

*Caleb Stamper[1,2], David Cortie[1,2], Zengji Yue,[2] Xiaolin Wang,[2] Dehong Yu[1]\**

[1]Australian Nuclear Science and Technology Organisation, Locked Bag 2001, Kirrawee DC NSW 2232, Australia.

[2]Institute for Superconducting and Innovative Materials, University of Wollongong, Wollongong, Australia.

**\*Corresponding Author**

Email: dyu@ansto.gov.au

An analytical model describing the vibrational phonon density of states (VDOS) of liquids has long been elusive, mainly due to the difficulty in dealing with the imaginary modes dominant in the low-energy region, as described by the instantaneous normal mode (INM) approach. Nevertheless, Zaccone and Baggioli have recently developed such a model based on overdamped Langevin liquid dynamics. The model was proposed to be the universal law for the vibrational density of states of liquids. Distinct from the Debye law, $g(\omega) \propto \omega^2$, for solids, the universal law for liquids reveals a linear relationship, $g(\omega) \propto \omega$, in the low-energy region. The universal law has been successfully verified with computer simulated VDOS for Lennard-Jones liquids. We further confirm this universal law with experimental VDOS measured by inelastic neutron scattering on real liquid systems including water, liquid metal, and polymer liquids. We have applied this model and extracted the effective relaxation rate for the short time dynamics for each liquid. The model has been further evaluated in the predication of the specific heat. The results have been compared with the existing experimental data as well as with values obtained by different approaches.

**TOC GRAPHICS**

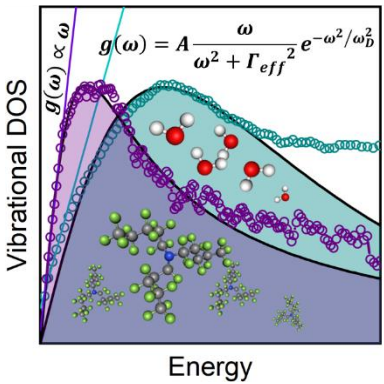

The vibrational density of states (VDOS, denoted $g(\omega)$) is a fundamental property of solid materials, determining their specific heat and thermal transport. For over 100 years, the Debye model has been the foundation for our understanding of the VDOS of solid materials which show a low energy relationship of $g(\omega) \propto \omega^2$, where ω is the frequency and g(ω) is the number of modes within an energy/frequency interval.[1] However, a corresponding model for the VDOS of liquids has been particularly elusive. The difficulty in defining the VDOS of liquids arises from their vastly more complex dynamics. While solids have been shown to be reasonably described by harmonic normal modes (i.e., acoustic phonons) at low temperatures, the strong intrinsic anharmonic dynamics of liquids makes applying a normal mode picture more challenging. Nonetheless, the normal mode picture has successfully been extended from solids to liquids in the form of instantaneous normal modes (INMs); defined by diagonalizing the Hessian or dynamical matrix of the potential energy at every instantaneous configuration.[2] The resulting modes consist of both real and imaginary frequencies, corresponding to stable and unstable modes respectively. The imaginary frequency modes reflect the instability of the time-frozen instant structure and arise from the saddle points in the potential energy landscape. They contain information about potential barrier hopping rates and cause the damping of the stable modes.[3] The presence of these imaginary modes makes it particularly challenging to develop an analytical expression of the VDOS of liquids. Recently, however, Zaccone and Baggioli have successfully derived such an analytical expression that is considered to be the universal law for the vibrational density of states of liquids.[2] Based on the overdamped Langevin dynamics, this universal law has defined the liquid VDOS in terms of the imaginary INMs, which have been shown to dominate the dynamics of liquids, particularly in the low energy region.[3-6] According to Zaccone and Baggioli (who we will refer to herein as Z-B) the VDOS for a liquid is given by the expression:[2]

$$g(\omega)_{total} = \frac{1}{3\pi N} \sum_j \frac{\omega}{\omega^2 + \Gamma_j^2}, \quad (1)$$

where $N$ is the number of atoms and $\Gamma$ is the relaxation rate of the $j^{th}$ INM mode. In the low energy limit, $\omega \ll \Gamma_j$, a linear relationship is revealed for the VDOS.

$$g(\omega) = \frac{1}{3\pi N} \sum_j \frac{1}{\Gamma_j^2} \omega \quad (2)$$

The sum over all $\Gamma_j$ can be replaced with an effective rate, $\Gamma_{eff}$, which is dominated by the lowest relaxation rate value (i.e., the longest-lived hydrodynamic mode).[2] This linear scaling law of $g(\omega) \propto \omega$ provides a signature of the low energy dynamics for liquids, analogous to the Debye square relationship for solid materials.

Equation 1 has been successfully tested against numerical simulated data on model liquids (e.g., Ar) based on the Lennard-Jones potential. However, to our knowledge, this universal law has not been confirmed experimentally on more complex liquid systems. For this purpose, we measured the VDOS for several liquids by inelastic neutron scattering (INS). We chose three classes of liquids: liquid gallium (Ga), water ($H_2O$), Fomblin oil ($CF_3O[-CF(CF_3)CF_2O-]_x(-CF_2O-)_yCF_3$) and Fluorinert liquid ($C_{15}F_{33}N$); from atomic, small molecular, to perfluorocarbon polymer type liquids.

The inelastic neutron scattering measurements are carried out on Pelican – the time-of-flight cold neutron spectrometer at the Australian Nuclear Science and Technology Organisation. The VDOS is determined through the experimental dynamic scattering function $S(Q, \omega)$ which is a function of momentum $Q$ and energy ($\hbar\omega$) transfer. Based on the incoherent one-phonon approximation, the VDOS for a Bravais powder sample (isotropic system) is given by:[7]

$$g(\omega) = C \int \frac{\omega}{Q^2} S(Q, \omega)\left(1 - e^{-\hbar\omega/k_B T}\right) dQ, \qquad (3)$$

where $C$ is a factor containing the atomic mass and Debye-Waller factor, $\exp(-2W)$, which is taken as unity for all samples here, $k_B$ is the Boltzmann constant, and $S(Q, \omega)$ is the dynamic scattering function of the sample. The integration over $Q$ covers the experiment-accessible range of momentum transfer.

For a non-Bravais sample, the measured phonon density of states (VDOS) is given by the neutron-weighted phonon density of states:

$$g_{NW}(\omega) = \sum_i f_i \frac{\sigma_i}{M_i} g_i(\omega), \qquad (4)$$

where the sum over $i$ includes all elements in the sample, $f_i$ is the $i^{\text{th}}$ atomic concentration, $\sigma_i$ is the total neutron bound cross section as both coherent and incoherent phonon scattering processes are measured experimentally, $M_i$ is the atomic mass, and $g_i(\omega)$ is the partial VDOS of the element $i$. According to Equation 4, for water, the measured VDOS is dominated by hydrogen as the $\sigma/M$ ratio for H (82) is much larger than O (0.26). For the two perfluorocarbon samples, the fluorine is the main contributor to the VDOS due to its large concentration as compared to other elements.

Figure 1a-d shows the VDOS measured by INS for the four liquid samples at temperatures of 340 K for Ga and 300 K for water, Fomblin, and Fluorinert. The Z-B model with a high-frequency Debye cut-off, $\omega_D$, is used in the following form to fit the experimental VDOS (based on Equation 1):[2]

$$g(\omega) = A \frac{\omega}{\omega^2 + \Gamma_{eff}^2} e^{-\omega^2/\omega_D^2}, \qquad (5)$$

where $A$ and $\Gamma_{eff}$ are treated as fitting parameters and $\omega_D$ is taken from the literature (see Table 1). For all the samples, a single low energy peak below 10 meV is observed. This is followed by some high-energy modes which include additional peaks that differ in each sample.

For water, as shown in Figure 1a, the initial peak is at ~6.5 meV. It has been suggested that this peak is dominated by collective modes of hydrogen-bond bending, perpendicular to the hydrogen bond (--) itself (O–H--O).[8] The large, broad peak centered around ~60 meV, meanwhile, is suggested to arise primarily due to librational motion caused by intermolecular coupling.[9] In between the two major peaks, around 20-35 meV, weak hydrogen bond stretching modes in line with the hydrogen bond (O-H--O) are expected to contribute to the VDOS.[10, 11] However, it has been shown by simulations based on the INM approach that the 6.5 meV peak corresponds to translational motions, while the 60 meV feature corresponds to the rotational motions around each of the three molecular axes.[6] These different paradigms (molecular vibrational normal modes and the INM approach) provide an understanding of the water VDOS from different angles. The first peak can be well fit by Equation 5 (Figure 1a), indicating that the Z-B model describes the low energy translational modes well. Interestingly, the aforementioned INM calculation showed that the imaginary/unstable modes contribute only 6% of the proportion of total modes – and were almost entirely translational in nature – compared to 10-30% seen in other simple liquids.[6] Since there is only a small contribution from the imaginary modes, the low energy dynamics of water may be governed by both the imaginary modes and the associated overdamped real modes.[3] It is not surprising that the Z-B model cannot model the stretching and rotational modes above ~20 meV as the theory was developed only for low energy imaginary modes.[2]

The hydrodynamics of liquid gallium have been of particular interest for quite some time now, with various contradictory pictures portrayed.[12-15] Interestingly, transverse acoustic phonon modes have been reported in liquid Ga near room temperature, despite liquid Ga being predominantly described by a "hard-sphere" model.[16] The acoustic transverse and longitudinal phonon-like modes are expected to appear in the VDOS at ~6 meV and ~18 meV respectively.[17] For liquid gallium, the observed first low energy peak of the VDOS (Figure 1b) can also be fit well with Equation 5 up to ~10 meV, after which the Z-B model consistently underestimates the experimental data until high energies where the theory and experiment converge. From the quality of the fitting, the low energy dynamics of liquid Ga are well-captured by INMs. This is consistent with reported INM simulations of liquid Ga.[3] This raises a question of how big the contribution is from acoustic transverse phonons to the low energy dynamics of liquid Ga and whether the ~6 meV feature is simply the manifestation of INMs.

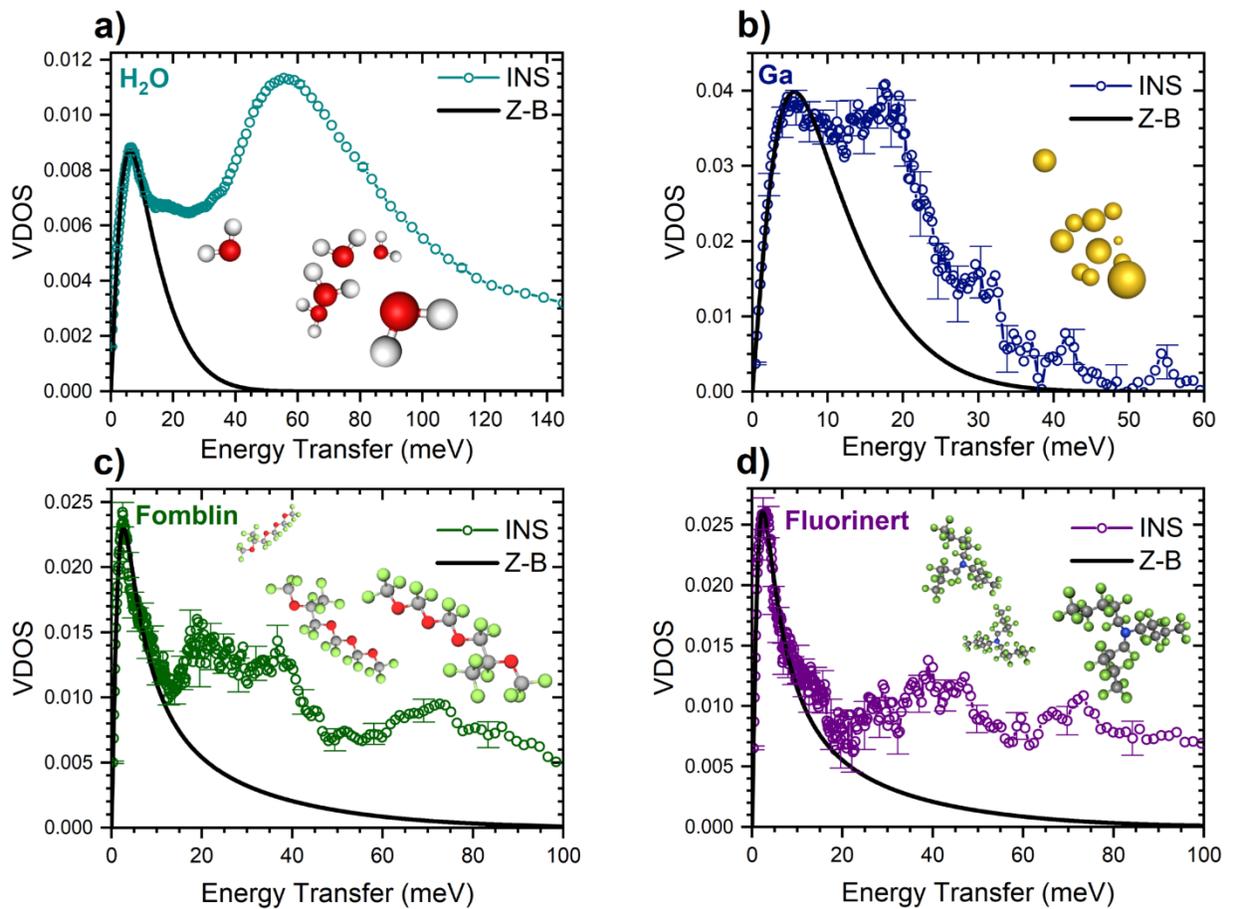

**Figure 1.** The VDOS obtained by INS (circles) for a) water, b) liquid gallium, c) Fomblin oil, and d) Fluorinert liquid. Each INS data set is fit with Equation 5 up to 10 meV (see Table 1 for the parameters used). Each figure has an inset schematic representing the atoms/molecules of the corresponding liquid. The atoms depicted in the ball-stick figures are oxygen (red), hydrogen

(white), gallium (yellow), carbon (grey), fluorine (green), and nitrogen (blue). Note that the depiction of Fomblin oil is of the form $CF_3O[-CF(CF_3)CF_2O-]_x(-CF_2O-)_yCF_3$ where $x = y = 1$ for illustrative purposes; the real sample has (on average) $x \approx 8$, $y \approx 28$.

The VDOS of both perfluorocarbons, in the solid state (at 21 K) have been studied by both INS and density functional theory in the context of understanding the neutron loss processes from the wall of the container for ultra-cold neutron storage used for neutron lifetime measurements.[18, 19] From this study, the low energy VDOS contribution was primarily assigned to torsions of the fluoromethyl (O–CF$_3$) groups at the end of the chain, and in-chain C–C torsion, whereas at ~25-45 meV, deformational in-plane modes (denoted δ[O–C–C], δ[O–C–F], and δ[C–O–C]) are responsible for additional VDOS.[19] These characteristics are clearly present in the current measurements on both Fomblin and Fluorinert liquid samples which look strikingly similar to the solid state VDOS of these materials at 21 K – likely due to the similarity of the disordered nature of the complex solid to the liquid phase. The VDOS of Fomblin oil and Fluorinert liquid (Figure 1c and 1d, respectively) are both very well represented by the Z-B model up to ~15-20 meV. This tells us that the low energy torsional modes in the solid state may change to INMs in the liquid state due to the change in the potential energy landscape.

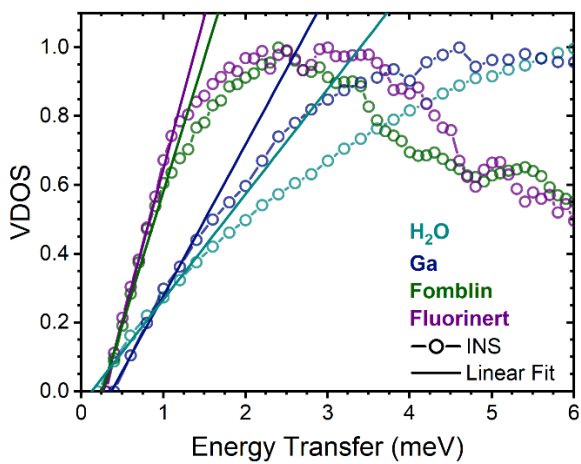

**Figure 2.** Low energy VDOS of the four different liquids obtained by INS (circles); normalized to their respective local peaks for clarity. The INS data is fit with Equation 6 up to 1 meV (solid lines) before being normalized. The error bars are not displayed for clarity.

**Table 1.** Values used during the fitting procedure whereby Equations 5 and 6 were used to fit the INS data in Figure 1 and 2, respectively. Equivalent Debye temperature values for the Debye cut-off frequency, $\omega_D$ are also indicated in square brackets. The value of $\omega_D$ for Fluorinert is assumed to be the same as Fomblin. The presented errors are determined from the fitting procedure.

|           | Full model fit (Eq. 5) |                       |                                     | Low-energy linear fit (Eq. 6)       |
|-----------|------------------------|-----------------------|-------------------------------------|-------------------------------------|
| Liquid    | A* (dimensionless)     | $\omega_D$ (meV) [K]  | $\Gamma_{eff}$* (meV) [ps$^{-1}$]   | $\Gamma_{eff}$* (meV) [ps$^{-1}$]   |
| H$_2$O    | 0.130 ± 0.004          | 24                    | 6.92 ± 0.17                         | 7.3 ± 0.4                           |
|           |                        | [280] [20]            | [10.5 ± 0.1]                        | [11.0 ± 0.7]                        |
| Ga        | 0.563 ± 0.015          | 20.0                  | 6.47 ± 0.14                         | 5.66 ± 0.07                         |
|           |                        | [240] [21]            | [9.8 ± 0.2]                         | [8.66 ± 0.07]                       |
| Fomblin   | 0.121 ± 0.002          | 64.6                  | 2.62 ± 0.05                         | 2.68 ± 0.07                         |
|           |                        | [750] [19]            | [3.98 ± 0.08]                       | [4.1 ± 0.1]                         |
| Fluorinert| 0.123 ± 0.001          | 64.6                  | 2.36 ± 0.03                         | 2.60 ± 0.04                         |
|           |                        | [750]                 | [3.59 ± 0.06]                       | [3.95 ± 0.07]                       |

*Denotes a fitting parameter.

Figure 2 shows the low energy VDOS of the four liquids, fit with a simple linear expression (6) based on Equation 2 (note that all the data in Figure 2 is normalized for comparative purposes, the linear fits were obtained by fitting the data in Figure 1):

$$g(\omega) = A \frac{1}{\Gamma_{eff}^2} \omega \tag{6}$$

The parameter $A$ is the same parameter obtained in the fitting of Equation 5. The VDOS for each liquid shows strong linearity up to ~1-1.5 meV, as shown in Figure 2. This clearly confirms the universal law of the vibrational density of states for liquid derived by Zaccone and Baggioli recently.[2] This linear relation of the low energy VDOS has been predicted by several simulations for other liquids and some disordered materials.[22-24]

The effective relaxation rates, $\Gamma_{eff}$, corresponding to the first peak position of the VDOS given by the Z-B model (Equation 1), are given in Table 1. For $H_2O$, the $\Gamma_{eff}$ value (10.5 ps$^{-1}$) is slightly faster than the relaxation time of the largest hydrogen bonding configuration ($\tau \approx 0.15$ ps → $\Gamma \approx 6.67$ ps$^{-1}$) reported in the literature (in an instant, the larger the fraction of hydrogen bonding present for a given configuration, the longer its lifetime).[25, 26] Hence, the effective relaxation time from fitting the Z-B model is dominated by modes much faster than the diffusive dynamics (around 30 ps).[27] For liquid Ga, two short-time diffusive processes with different lifetimes have been identified by molecular dynamics simulations and quasi-elastic neutron scattering: a slow self-diffusion with $\tau \approx 0.9$ ps ($\Gamma \approx 1.1$ ps$^{-1}$), and a faster cage effect whereby a gallium atom is trapped in a cage from its nearest neighbors for a time, $\tau \approx 0.2$ ps ($\Gamma \approx 5$ ps$^{-1}$), before being able to diffuse.[28, 29] The fit of the Z-B model to the current experimental low energy liquid Ga VDOS gives $\Gamma_{eff} = 9.8$ ps$^{-1}$, faster than that of the relaxation process due to the cage effect.[29] For Fomblin and Fluorinert, based on the Z-B model, we derived the short time dynamics corresponding to $\Gamma_{eff} = 2.62$ ps$^{-1}$ and $\Gamma_{eff} = 2.36$ ps$^{-1}$ respectively. Their slower relaxation rate in the short-time scale compared to water and gallium is intuitive, considering the much larger molecular liquids with average molecular weight of 3300 amu and 820 amu, respectively.

The linear fitting of the low energy region of the VDOS using Equation 6 – fixing the pre-factor to the values obtained by the fitting with Equation 5 – yielded very similar values for the effective relaxation rate compared to the full fitting with Equation 5 (Table 1). This highlights the linearity of the low energy VDOS of liquids and that the slope itself – even up to just 1 meV – contains the same information about the average lifetime of the INMs as defined by the peak position of the Z-B model. It is interesting to note that in all cases the linearity occurs consistently up to ~ 1 meV, despite the first peak in the VDOS occurring at different energies. This has also been observed from the simulation on Lennard-Jones liquids.[2] Such a result shows universal low energy hydrodynamic behavior amongst varied liquids.

Under the approximation of Equation 3, at low energies, we should have

$$g(\omega) \sim \omega^2 S(Q,\omega) \qquad (7)$$

Since $g(\omega)$ has a linear relationship with $\omega$, according to Z-B (Equation 2), we expect that $S(Q,\omega)$ should then behave as $1/\omega$. This would result in a similar or larger contribution compared to the $1/\omega^2$ behavior from a Lorentzian function representing the quasielastic neutron scattering (QENS) due to diffusive motions. This observation implies that the well accepted fitting of QENS spectra with Lorentzian functions and a constant background may need to be revisited in order to consider the contribution from the INMs of liquids.

Zaccone and Baggioli have recently applied the universal law for liquid VDOS to analytically calculate $c_V$ for several liquids, including noble elements and $N_2$.[30] The specific heat is given by the following formula:

$$c_V = k_B \int_0^\infty \left(\frac{\hbar\omega}{2k_BT}\right)^2 \sinh\left(\frac{\hbar\omega}{2k_BT}\right)^{-2} g(\omega)d\omega, \qquad (8)$$

where the symbols have their usual meaning. The monotonic decreasing of the specific heat with increasing temperature – because of Arrhenius-type relaxation of INMs – derived from Equation 8 with the VDOS given by Equation 1 agrees very well with the experimental data of those Lennard-Jones liquids. How well does this apply to other more complex liquid systems? In the following we shall calculate the specific heat based on the Equation 8 with the experimental VDOS and that given by Z-B model for the four liquid systems studied here.

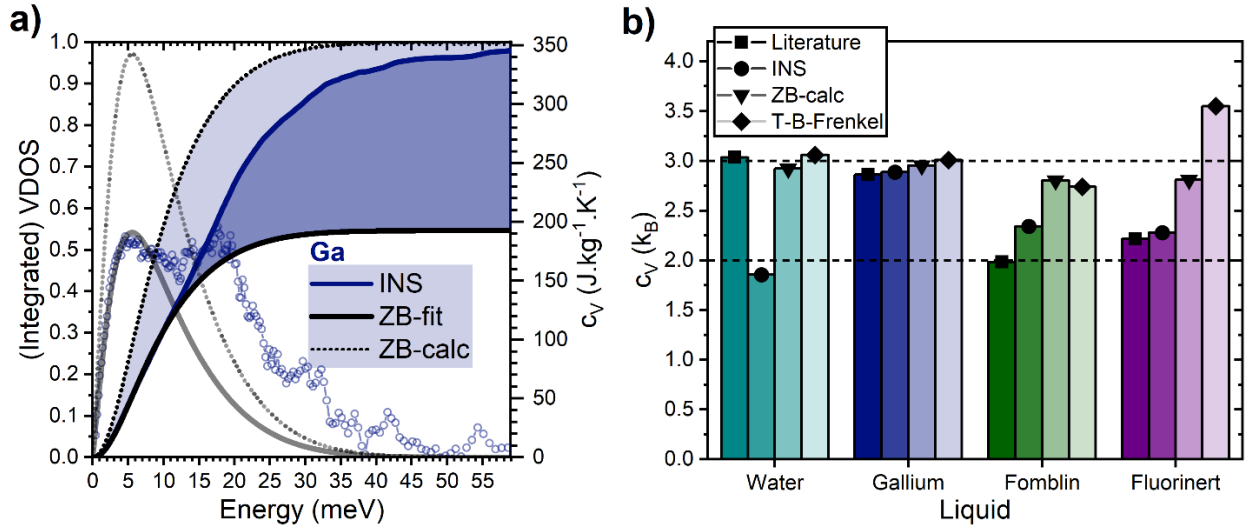

**Figure 4.** a) The cumulative specific heat, $c_V$, (right axis) of gallium, as a function of energy as calculated from the experimental VDOS (blue), the Z-B fitting (black solid line) and the renormalized Z-B VDOS (black dotted line). The corresponding VDOS (empty circles, faded black solid line, and faded black dotted line, respectively) is provided for reference (left axis),

and b) the specific heat (in units of $k_B$) for each liquid as per the literature, derived from the experimental VDOS, ZB-calc (dotted line in a)), and T-B-Frenkel calculation.

Taking liquid Ga as an example, the value of $c_V = 340 \pm 50$ J.kg$^{-1}$.K$^{-1}$ at 340 K has been obtained numerically with the experimental VDOS using Equation 8. It agrees remarkably well with the measured value reported in the literature, $c_V = 341$ J.kg$^{-1}$.K$^{-1}$ at 350 K.[31] It is obvious that the fit of the Z-B model only catches the low energy peak of the VDOS. Thus, the direct use of the Z-B fit VDOS results in a significant underestimation (~60%) of the specific heat (as ZB-fit). We take an alternate approach in which we keep the fitted value of $\Gamma_{eff}$ and renormalize the Z-B VDOS to unity (equivalent to fixing $A = 1/3\pi N$). By doing this we essentially compress the high energy modes into the low energy region represented by INMs. Using the renormalized Z-B VDOS, the $c_V$ (as ZB-calc) is 352 J.kg$^{-1}$.K$^{-1}$, only slightly larger than the value calculated from the experimental VDOS, despite the low-energy modes being over-represented. The comparison of these approaches is shown in Figure 4a with the cumulative specific heat versus energy. It clearly indicates that the ZB-calc overestimates the result in the low energy region, while the ZB-fit underestimates the high energy region significantly, with respect to the result from the experimental VDOS. The same analysis is undertaken for the other liquids and the resulting values for $c_V$ are presented in Table 2.

**Table 2.** Values of specific heat, $c_V$, for each liquid as calculated by using experimental VDOS (labeled as INS), ZB-calc, T-B-Frenkel, and the existing experimental values from previous studies in the literature (listed as Literature). Errors are estimated from the statistic uncertainty of the experimental VDOS and the resulting uncertainty in the fitted relaxation rates.

| Liquid | H$_2$O $c_V$ (J.kg$^{-1}$.K$^{-1}$) | Ga $c_V$ (J.kg$^{-1}$.K$^{-1}$) | Fomblin $c_V$ (J.kg$^{-1}$.K$^{-1}$) | Fluorinert $c_V$ (J.kg$^{-1}$.K$^{-1}$) |
|---|---|---|---|---|

|              | at 300 K      | at 340 K    | at 300 K      | at 300 K      |
|--------------|---------------|-------------|---------------|---------------|
| Literature   | 4130 [32]     | 341 [31]    | 1000 [33]     | 1100 [34]     |
| INS          | 2570 ± 30     | 340 ± 50    | 1190 ± 110    | 1130 ± 150    |
| ZB-calc      | 4050 ± 100    | 352 ± 8     | 1420 ± 30     | 1400 ± 20     |
| T-B-Frenkel  | 4240 ± 100    | 359 ± 8     | 1390 ± 30     | 1760 ± 20     |

As a further comparison, we evaluate the specific heat with an alternate approach, as described by Trachenko and Brazhkin,[35] based on a hybrid hydrodynamic-solid description of liquids, originally developed by Frenkel.[36, 37] In this picture, a frequency, denoted as the Frenkel frequency $\omega_F = 1/\tau$, is defined, where $\tau$ is the dominant diffusive-jump time. Firstly, the dynamics at low frequencies ($< \omega_F$) (timescales larger than $\tau$) are considered to be hydrodynamic in nature, behaving diffusively and sustaining one longitudinal phonon mode, but not shear stress (transverse modes). Secondly, at high frequencies ($> \omega_F$) (timescales significantly less than $\tau$), there is no rearrangement of particles, and the system is considered to be a solid glass, supporting two transverse phonon modes and one longitudinal phonon mode. Accordingly, an analytical expression for energy, $E$, and the specific heat of liquids (in units of $k_B$), which we will herein refer to as the T-B-Frenkel model, are given below:[35]

$$E = NT\left(1 + \frac{\alpha T}{2}\right)\left(3D\left(\frac{\hbar\omega_D}{T}\right) - \left(\frac{\hbar\omega_F}{\omega_D}\right)^3 D\left(\frac{\hbar\omega_F}{T}\right)\right), \quad (9)$$

$$c_V = \frac{1}{N}\frac{dE}{dT},$$

where, $\alpha$ is the coefficient of thermal expansion (taken from the literature),[34, 38-40] $D(x) = 3/x^3 \int_0^x z^3 dz/\exp(z) - 1$ is the Debye function, and the other symbols have their usual meaning. We determine $\omega_F$ through $\Gamma = \hbar\omega_F$, where the relaxation rates correspond to the effective rate obtained from the fitting of Z-B model to the experimental VDOS. The $c_V$ calculated using Equation 9 for each liquid is listed in Table 2 (T-B-Frenkel) and presented in Figure 4b (in units of $k_B$). Typically, the $c_V$ of liquids ranges from ~$3k_B$ at low temperature to ~$2k_B$ at high temperature. This loss of $c_V$ as a function of temperature is explained in the Frenkel framework by the loss of transverse phonon modes above $\omega_F$ with increasing temperature.[35] As shown in Figure 4b, for water, the $c_V$ values from both the Z-B model and the T-B Frenkel model match very well with the literature value. However, the value from the experimental VDOS is only about half of the values derived with other approaches. This is because the experimental VDOS for water is skewed heavily by the large neutron scattering cross section of hydrogen compared to oxygen, and hence significantly underestimates the lower energy modes which provides a dominate contribution to $c_V$. This effect is less pronounced for Fomblin and Fluorinert, though with fluorine dominating the signal for both samples, and the $c_V$ derived from the experimental VDOS matches well with the literature values. The ZB-calc and T-B-Frenkel models both do a relatively good job at predicting the $c_V$ of the perfluorocarbons, though both overestimate it. The T-B-Frenkel value for Fluorinert is an outlier and requires further investigation. It is noted that a lower value $c_V$ = 1350 J.kg$^{-1}$.K$^{-1}$ = $2.7k_B$ is obtained if the term for the thermal expansion is dropped in Equation 9. For Ga, similar to water, all approaches give a close value to the literature value ~$3k_B$, providing strong evidence of a solid-like liquid state and consistent with the observation of dispersive, transverse phonon-like modes in liquid Ga.[16]

To conclude, we have provided the first experimental confirmation of the recently derived universal law for the vibrational density of states of liquids by Zaccone and Baggioli based on the INM theoretical approach. The VDOS measured by inelastic neutron scattering were fit extremely well up to ~10-15 meV by the Z-B model for four different samples of H$_2$O, Ga metal, and perfluorocarbon liquids. The universal linear relationship of the VDOS against energy predicted by the Z-B model was validated experimentally for all samples studied. The effective short-time dynamic relaxational rates of each liquid were derived from the fit of the Z-B model to the experimental VDOS. Compared to simulations in the literature, the effective relaxation rates determined here – assumed to approximately represent the longest-lived hydrodynamic mode – are faster than that for the diffusive dynamics. The application of the Z-B universal law to the prediction of the specific heat for complex liquid systems has been evaluated and compared with available experimental data as well as with a different theoretical model. It is shown that the Z-B model can reasonably reproduce the experimental specific heat with a slight overestimation due to overrepresentation of the high frequency modes as low energy INMs. It is observed from the linear relationship of the low energy VDOS that the neutron scattering function of a liquid should behave as $1/\omega$ rather than $1/\omega^2$ from a Lorentzian function representing the diffusive dynamics in this energy range. This may need to be considered in the analysis of quasielastic neutron spectra for liquids. We believe the work presented here will motivate further development in studying liquid dynamics both theoretically and experimentally.

Methodology

**INS measurements.** The inelastic neutron scattering (INS) experiments are conducted on the time-of-flight cold-neutron spectrometer, PELICAN, at the Australian Centre for Neutron Scattering, ANSTO[41]. The instrument is configured with an incident neutron wavelength of 4.69 Å, corresponding to an incident energy of 3.72 meV with an energy resolution of 0.135 meV at the elastic line. A mass of 0.5 g of general-purpose laboratory grade of deionized water is filled inside an annular aluminum can having 0.1 mm gap. Annular aluminum cans having 0.2 mm gap are filled with 2 g of commercial available Fomblin (YL 25/6, Fomblin® Y LVAC 25/6 | Solvay);[42] $CF_3O[-CF(CF_3)CF_2O-]_x(-CF_2O-)_yCF_3$, and Fluorinert (FC-70, Fluorinert | 3M);[34] $C_{15}F_{33}N$. The estimated transmission is from 90% to 99%. Approximately 3.2 g of Ga (Gallium | RotoMetals)[43] is inserted into a 4 mm diameter Teflon tube which was placed inside an aluminum cylinder sample can with a 25 mm diameter. Transmission of 80% is estimated for a sample with thickness of 4 mm. The INS measurements were carried out at 300 K except for Ga, which was performed out at 340 K. The corresponding empty can is measured in the same conditions as for the samples for background subtraction. In addition, a standard cylinder shape vanadium sample is also measured for detector efficiency normalization and energy resolution calibration. The data reduction, including background subtraction and detector normalization are performed using the Large Array Manipulation Program (LAMP)[44]. The phonon density of states was obtained on the neutron energy gain side with normalization to the total area and no multiphonon scatterings are corrected. The range of momentum transfer covered is from 0.2 Å to 7 Å depending on the range of energy transfer considered.

**Data Fitting.** Fitting was done using OriginLab, OriginPro 2022.

AUTHOR INFORMATION

Caleb Stamper – https://orcid.org/0000-0002-4392-2369
David Cortie – https://orcid.org/0000-0003-2383-1619
Zengji Yue – https://orcid.org/0000-0001-5917-5057
Xiaolin Wang – https://orcid.org/0000-0003-4150-0848
Dehong Yu – https://orcid.org/0000-0003-2995-0336

**Notes**

The authors declare no competing financial interests.

ACKNOWLEDGMENTS

The authors acknowledge beam time awarded by ANSTO (proposal DB9480). C.S. acknowledges support by the Australian Government Research Training Program through the University of Wollongong.